\documentclass[a4paper,twocolumn,english]{acm_proc_article-sp}
\usepackage[T1]{fontenc}
\usepackage[latin9]{inputenc}
\usepackage{babel}
\usepackage{textcomp}
\usepackage{amsmath}
\usepackage{graphicx}
\usepackage[unicode=true,
 bookmarks=true,bookmarksnumbered=false,bookmarksopen=false,
 breaklinks=false,pdfborder={0 0 1},backref=false,colorlinks=false]
 {hyperref}
\hypersetup{
 pdfauthor={Andrew N. Jackson}}

\makeatletter

\pdfpageheight\paperheight
\pdfpagewidth\paperwidth

\numberwithin{equation}{section}
\numberwithin{figure}{section}
\newenvironment{lyxlist}[1]
{\begin{list}{}
{\settowidth{\labelwidth}{#1}
 \setlength{\leftmargin}{\labelwidth}
 \addtolength{\leftmargin}{\labelsep}
 }}
{\end{list}}

\makeatother

\begin{document}
\author{ 
% e.g. one 'row of three' or two rows (consisting of one row of three % and a second row of one, two or three). 
% 1st. author 
\alignauthor
Andrew N. Jackson\\
       \affaddr{The British Library}\\
       \affaddr{Boston Spa, Wetherby}\\
       \affaddr{West Yorkshire, LS23 7BQ, UK}\\
       \email{Andrew.Jackson@bl.uk}
}
% Just remember to make sure that the TOTAL number of authors 
% is the number that will appear on the first page PLUS the 
% number that will appear in the \additionalauthors section.

\title{Formats over Time: Exploring UK Web History}
\maketitle
\begin{abstract}
Is software obsolescence a significant risk? To explore this issue,
we analysed a corpus of over 2.5 billion resources corresponding to
the UK Web domain, as crawled between 1996 and 2010. Using the DROID
and Apache Tika identification tools, we examined each resource and
captured the results as extended MIME types, embedding version, software
and hardware identifiers alongside the format information. The combined
results form a detailed temporal format profile of the corpus, which
we have made available as open data. We present the results of our
initial analysis of this dataset. We look at image, HTML and PDF resources
in some detail, showing how the usage of different formats, versions
and software implementations has changed over time. Furthermore, we
show that software obsolescence is rare on the web and uncover evidence
indicating that network effects act to stabilise formats against obsolescence. 

%A category including the fourth, optional field follows... 
\end{abstract}

\section{Introduction}

In order to ensure that our digital resources remain accessible over
time, we need to fully understand the software and hardware dependencies
required for playback and re-use. The relationship between bitstreams
and the software that makes them accessible is usually expressed in
terms of data `format' - instead of explicitly linking individual
resources to individual pieces of software, we attach identifiers
like file extensions, MIME types and PRONOM IDs to each and use that
to maintain the link. These identifiers can also be attached to formal
format specifications, if such documentation is available. 

Successful digital preservation therefore requires us to fully understand
the relationship between data, formats, software and documentation,
and how these things change over time. Critically, we must learn how
formats become obsolete, so that we might understand the warning signs,
choices and costs involved. This issue, and the arguments around the
threat of obsolescence, can be traced back to 1997, when Rothenburg
asserted that ``Digital Information Lasts Forever\textemdash{}Or
Five Years, Whichever Comes First.'' {[}1{]}. Fifteen years later,
Rothenberg maintains that this aphorism is still apt {[}2{]}. If true,
this implies that all formats should be considered brittle and transient,
and that frequent preservation actions will be required in order to
to keep our data usable. In contrast, Rosenthal maintains that this
is simply not the case, writing in 2010 that ``when challenged, proponents
of {[}format migration strategies{]} have failed to identify even
one format in wide use when Rothenberg {[}made that assertion{]} that
has gone obsolete in the intervening decade and a half.'' {[}3{]}.
Rosenthal argues that the network effects of data sharing act to inhibit
obsolescence and ensure forward migration options will arise. Similarly,
Rothenburg remains skeptical of the common belief that different types
of content are normalising on HTML5 and so reducing the number of
formats we need to address {[}2{]}. If these assertions are true,
then format migration or emulation strategies become largely unnecessary,
leaving us to concentrate on storing the content and simply making
use the available rendering software. 

The fact that the very existence of software obsolescence remains
hotly disputed therefore undermines our ability to plan for the future.
To find a way forward, we must examine the available evidence and
try to test these competing hypotheses. In this paper, we begin this
process by running identification tools over a suitable corpus, so
that we can use the resulting format profile to explore what happens
when formats are born, and when they fade away. Working in partnership
with JISC and the Internet Archive (IA), we have been able to secure
a copy of the IA web archives relating to the UK domain, and host
it on our computer cluster. The collection is composed of over 2.5
billion resources, crawled between 1996 and 2010, and thus gives us
a sufficiently long timeline over which some reasonable conclusions
about web formats might be drawn.

Determining the format of each resource is not easy, as the MIME type
supplied by the originating server is often malformed {[}4{]}. Instead,
we apply two format identification tools to the content of each resource
- DROID and Apache Tika. Both use internal file signature (or `magic
numbers') to identify the likely format of each bitstream, but differ
in coverage, complexity and granularity. In particular, DROID tuned
to determine different versions of formats, while Apache Tika returns
only the general format type, but augments it with more detailed information
gleaned from parsing the bitstream. Thus, by combining both sets of
results, we can come to a more complete understanding of the corpus.
Furthermore, by comparing the results from the different identification
tools, we can also uncover inconsistencies, problematic formats and
weak signatures, and so help drive the refinement of both tools.

\section{Method}

The test corpus is called the JISC UK Web Domain Dataset (1996-2010),
and contains over 2.5 billion resources harvested between 1996 and
2010 (with a few hundred resources dated from 1994), either hosted
on UK domains, or directly referenced from resources hosted under
`.uk'. This adds up to 35TB of compressed content held in 470,466
arc.gz and warc.gz files, now held on the a 50-node HDFS filesystem.
As the content is hosted on this distributed filesystem, we are able
to run a range of tools over the whole dataset in a reasonable time
using Hadoop's Map-Reduce framework.

Due to it's prominence among the preservation community and the fine-grained
identification of individual versions of formats, DROID was chosen
as one of the tools. To complement this, we also chose to use the
popular Apache Tika identification tool, which has been shown to have
much broader format coverage {[}5{]}. Unfortunately, both tools required
some modification in order to be used in this context. DROID was particularly
problematic, and we were unable to completely extract the container-based
identification system in a form that made it re-usable as a Map-Reduce
task. However, the binary file format identification engine could
be reused, and the vast majority of the formats that DROID can identify
are based on using that code (and the DROID signature file it depends
upon - we used signature file version 59). Herein, we refer to this
as the 'DROID-B' tool. Both tools were run directly on the bitstreams,
rather than being passed the URLs or responses in question, and so
the identification was based upon the resource content rather than
the name or any other metadata. For this first experimental scan,
we decided to limit the identification process to the top-level resource
associated with each URL and crawl time - archive or container formats
were not unpacked.

In order to compare the results from DROID-B and Apache Tika with
the MIME type supplied by the server, the identification results are
normalised in the form of extended MIME Types. That is, where we know
the version of a format as well as the overall MIME Type, we add that
information to the identifier using a standard type parameter, e.g.
``image/png; version=1.0'', corresponding to PUID fmt/11. In this
way, extended MIME types can act as a bridge between the world of
PRONOM identifiers and the standard identification system used on
the web. Broad agreement between tools can be captured by stripping
off the parameters, but their presence lets more detailed information
be collected and compared in simple standard form. A number of formats
also embed information about the particular software or hardware that
was using in their creation - PDF files have a `creator' and a `producer'
field, and many image formats have similar EXIF tags. As we are also
interested in the relationship between software and formats, we have
attempted to extract this data and embed it in the extended MIME type
as software and hardware parameters. The full identification process
also extracted the year each resource was crawled, and combines this
with the three different MIME types to form a single `key'. These
keys were then collected and the total number of resources calculated
for each. Overall, the analysis was remarkably quick, requiring just
over 24 hours of continuous cluster time.

\section{Results}

\subsection{The Format Profile Dataset}

The primary output of this work is the format profile dataset itself%
\footnote{To download the dataset, see \href{http://dx.doi.org/10.5259/ukwa.ds.2/fmt/1}{http://dx.doi.org/10.5259/ukwa.ds.2/fmt/1}.%
}. Each line of this dataset captures a particular combination of MIME
types (server, Apache Tika and DROID-B), for a particular year, and
indicates how many resources matched that combination. For example,
this line:

\texttt{image/png image/png image/png; version=1.0 2004 102}

means that in this dataset there were 102 resources, crawled in 2004,
that the server, Tika and DROID-B all agreed have the format `image/png',
with the latter also determining the format version to be `1.0'. Due
disagreements over MIME types and the number different hardware and
software identifiers the overall profile is rather large, containing
over 530,000 distinct combinations of types and year. Below, we document
some initial findings drawn from the data. However, there is much
more to be gleaned from this rich dataset, and we have made it available
under an open licence (CC0) in the hope that others will explore and
re-use it.

\begin{figure}
\includegraphics[width=1\columnwidth]{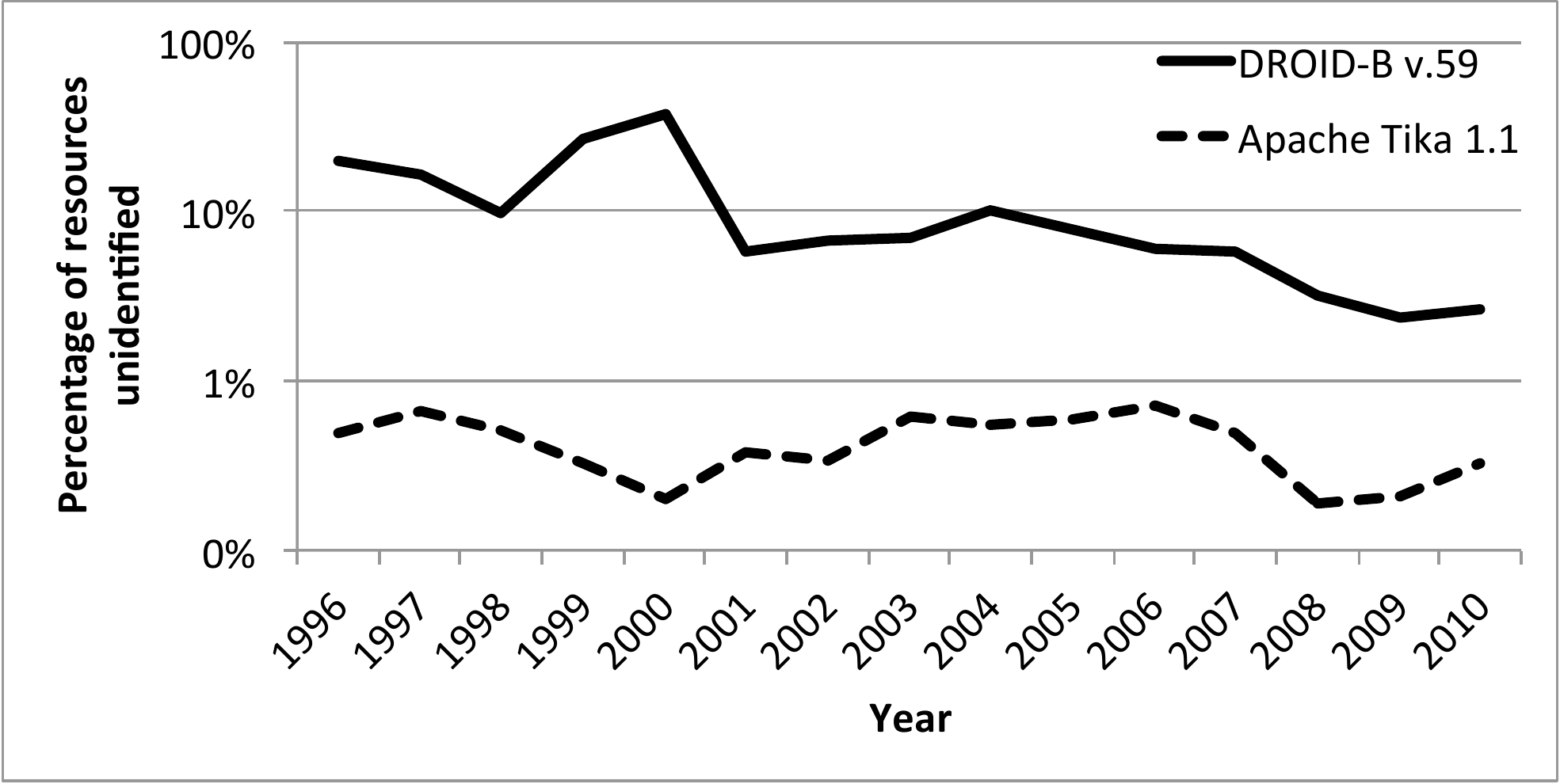}

\caption{\label{fig:Comparing-the-percentage}Identification failure rates
for Apache Tika and DROID-B.}
\end{figure}

\subsection{Comparing Identification Methods}

\subsubsection{Coverage \& Depth}

The identification failure rates for both tools are shown in Figure
\ref{fig:Comparing-the-percentage}, as a percentage of the total
number of resources from each year. Overall, Apache Tika has significantly
lower failure rate than DROID-B - 1\% versus around 10\%. There also
seems to be a significant downward trend in the DROID-B curve, which
would indicate that DROID copes less well with older formats. However,
initial exploration indicate that this is almost entirely due to the
prevalence of pre-2.0 HTML, which was often poorly formed.

\subsubsection{Inconsistencies}

By comparing the simple MIME types (no parameters) we were able to
compare the results from both tools, revealing 174 conflicting MIME
type combinations. For example, some 2,957,878 resources that Apache
Tika identified as `image/jpeg' we identified as `image/x-pict' by
DROID. The PRONOM signature for this format is rather weak (consisting
of a single byte value at a given offset) and can therefore produce
a large number of false positives when run at scale %
\footnote{Indeed, it appears that this signature has been removed from the latest
version of the DROID binary signature file (version 60, published
during preparation).%
}. Another notable class of weak signatures correspond to text-based
formats like CSS, JavaScript, and older or malformed HTML. Apache
Tika appears to perform slightly better here - for example, the HTML
signature is much more forgiving than the DROID-B signature.

More subtle inconsistencies arose for the Microsoft Office binary
formats and for PDF. In the former case, a full implementation of
DROID would probably be able to resolve many of the discrepancies.
The picture for PDF is more complex. The results were mostly consistent,
but DROID-B failed to recognise 1,340,462 resources that Apache Tika
identified as PDF. This appears to be because the corresponding PRONOM
signature requires the correct end-of-file marker (`\%\%EOF') to be
present, whereas many functional documents can be mildly malformed,
e.g. ending with `\%\%EO' instead. Also, the results for PDF/A-1a
and PDF/A-1b were not entirely consistent, with Tika failing to identify
many documents that DROID-B matched, but matching a small number of
PDF/A-1b documents that DROID missed. A detailed examination of the
signatures and software will be required to resolve these issues.

\subsection{Format Trends}

\begin{figure}
\includegraphics[width=1\columnwidth]{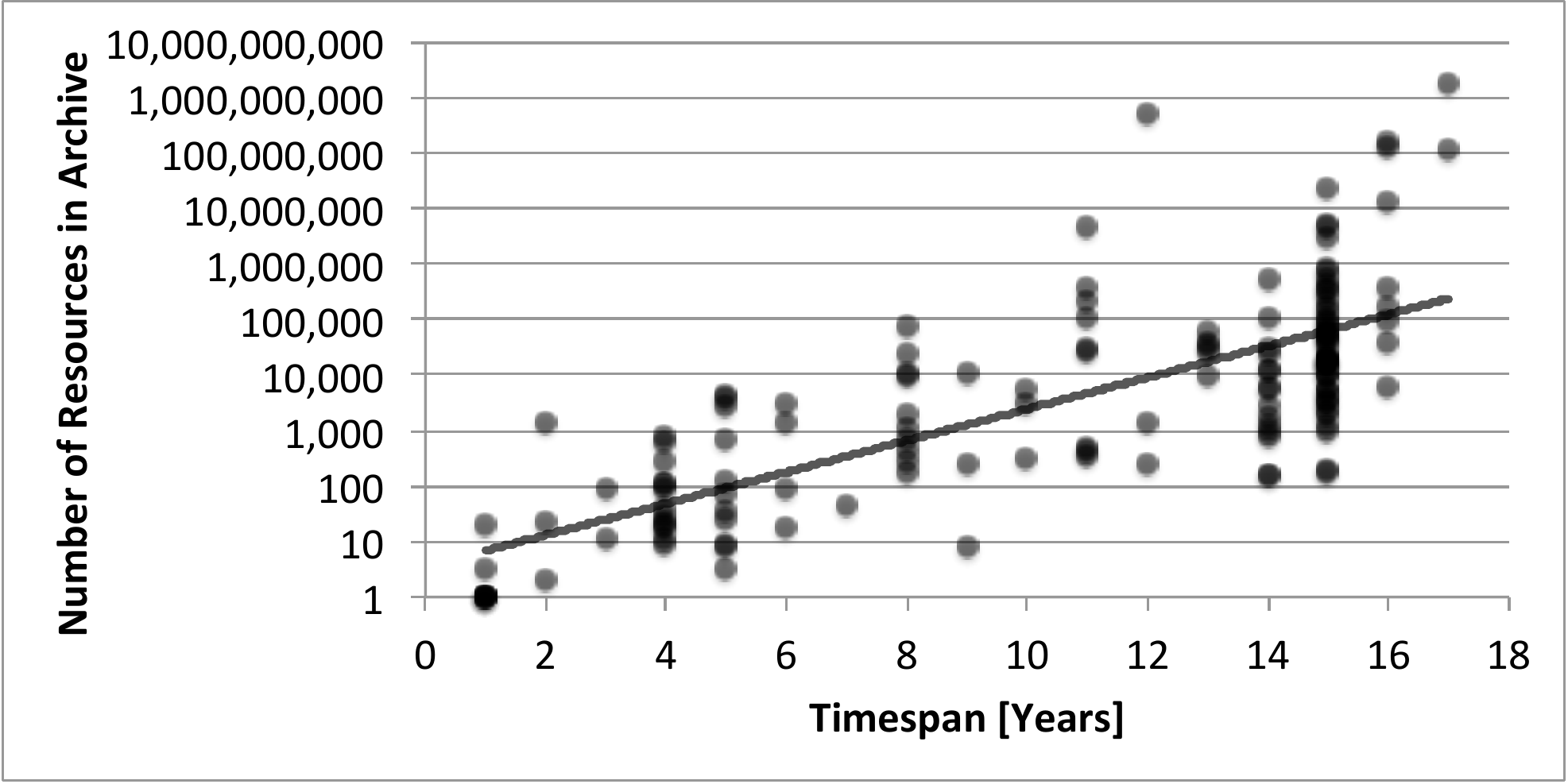}

\caption{\label{fig:Overall-results,-showing}Number of resources of each format
versus its lifespan. Formats identified using Apache Tika.}
\end{figure}

As mentioned in the introduction, one of the core questions we need
to understand is whether formats last a few years and then die off,
or whether (on the web at least) network effects take over and help
ensure formats survive. We start to examine this question by first
determining the lifespan of each format - i.e. the number of years
that have elapsed between a format's first and last appearance in
the archive. This lifespan is plotted against the number of resources
that were found to have that format, such that young and rare formats
appear in the bottom-left corner, whereas older and popular formats
appear in the top-right, as shown in figure \ref{fig:Overall-results,-showing}.
Due to the extreme variation in usage between formats, the results
are plotted on a logarithmic scale.

If popularity has no effect on lifespan, we would expect to see a
simple linear trend - i.e. a format that has existed for twice as
long as another would be found in twice as many documents. Due to
the logarithmic vertical axis of figure \ref{fig:Overall-results,-showing},
would be shown as a sharp initial increase followed by an apparent
plateau. However, in the presence of network effects we would expect
a much stronger relationship, and indeed this is what we find - a
format that has been around longer is exponentially more common that
younger formats (an exponential fit appears as a straight line in
figure \ref{fig:Overall-results,-showing}). A large number of formats
have persistent for a long time (47 formats have been around for 15
years), and that since 1997, roughly six new formats have appeared
each year while fewer have been lost (roughly 2 per year). While this
confirms the presence of the network effects Rosenthal proposed, proving
that these formats are more resilient against obsolescence will require
a deeper understanding of obsolescence itself.

\begin{figure}
\includegraphics[width=1\columnwidth]{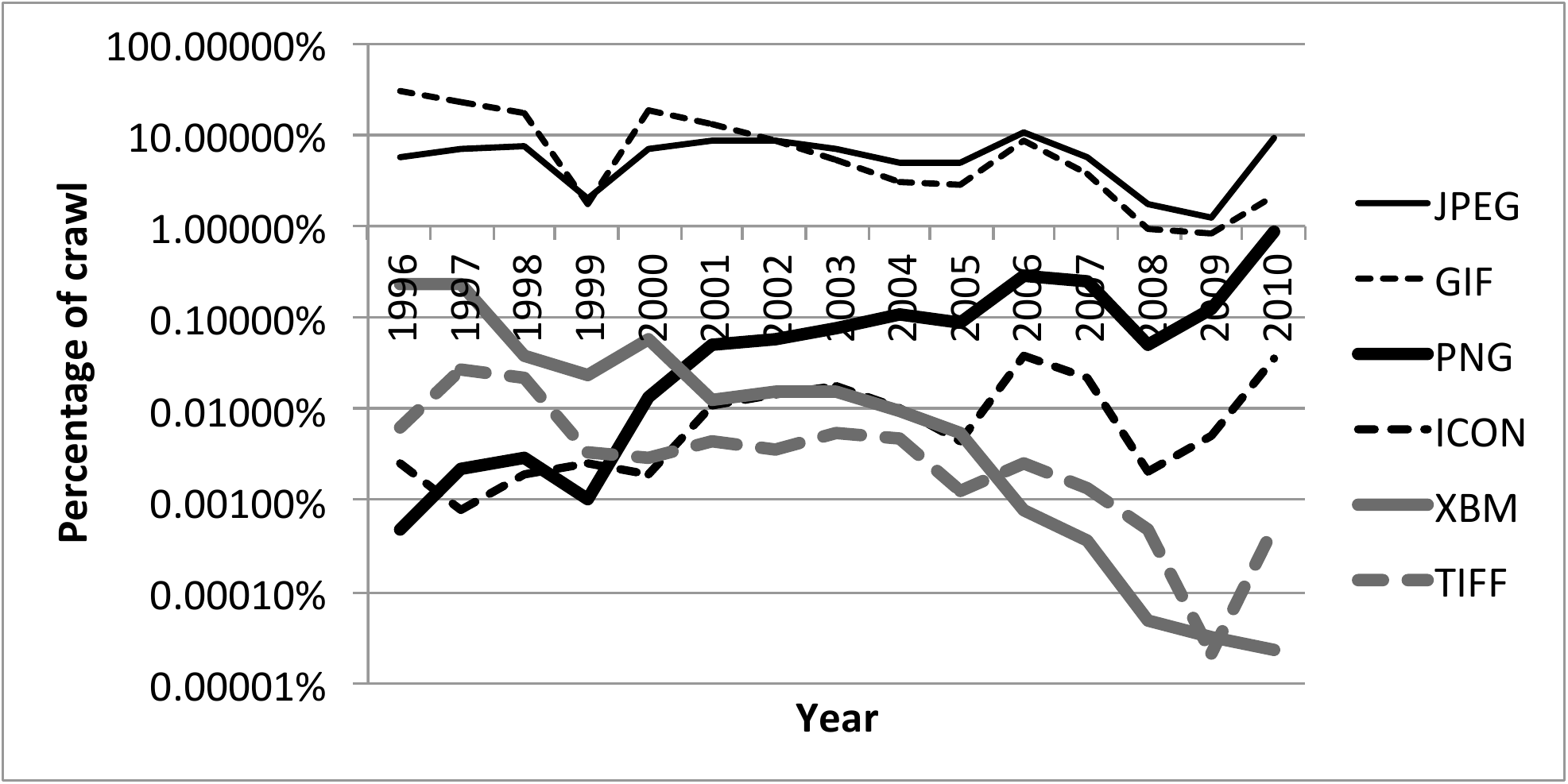}

\caption{\label{fig:Selected-trends-for}Selected popular image formats over
time. Formats identified using Apache Tika.}
\end{figure}

As a first step in that direction, we examine how format usage changes
over time. Figure \ref{fig:Selected-trends-for} shows the variation
in usage of some of the most common image formats. Unsurprisingly,
JPEG has remained consistently popular. In contrast, the PNG and ICO
formats have become more popular over time, and the GIF, TIFF and
XBM formats have decreased in popularity, with the drop in usage of
the XBM format being particularly striking.

\begin{figure}[t]
\includegraphics[width=1\columnwidth]{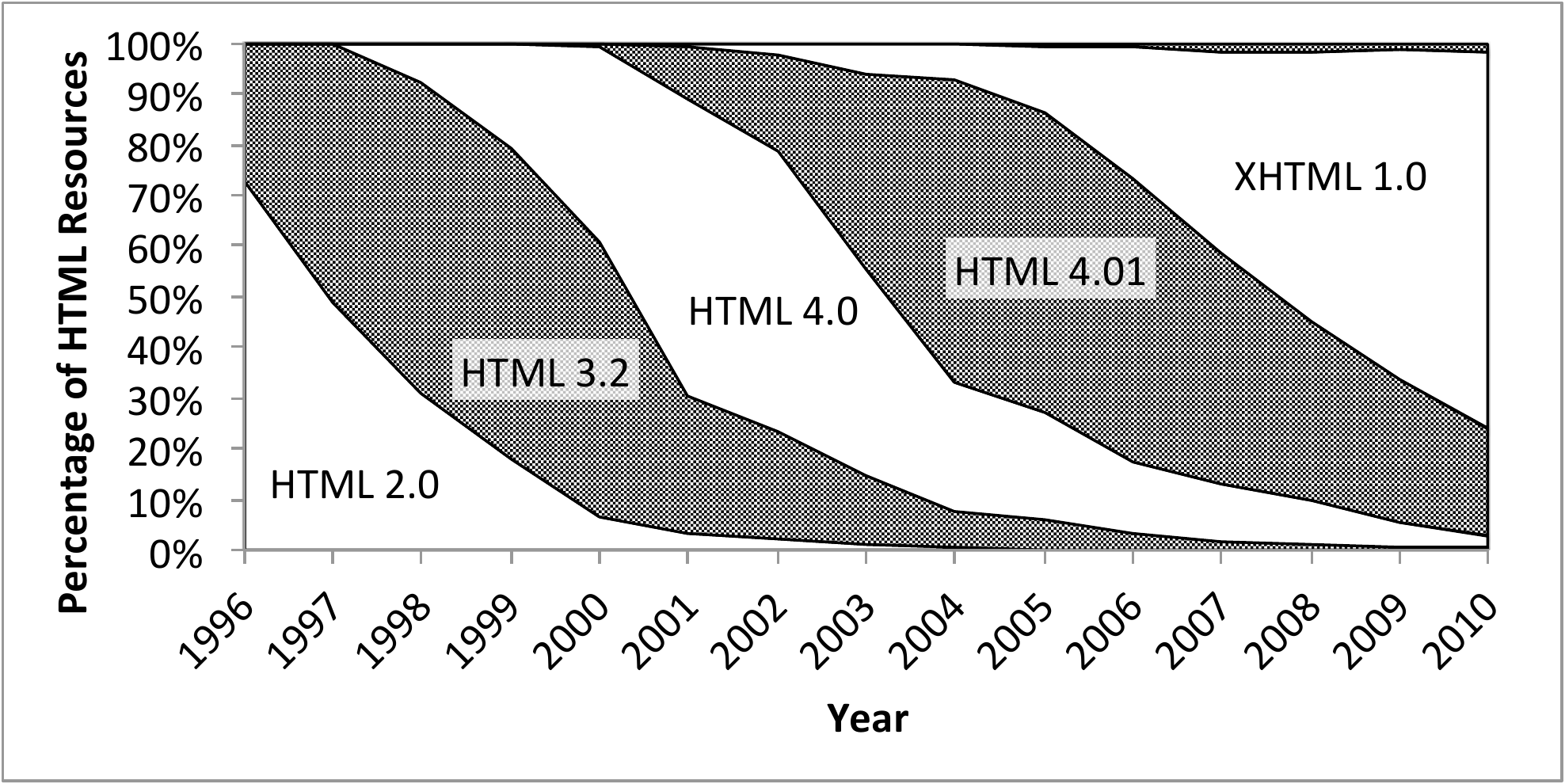}

\caption{\label{fig:Percentage-of-HTML}HTML versions over time. Formats identified
using DROID-B.}
\end{figure}

\begin{figure}[t]
\includegraphics[width=1\columnwidth]{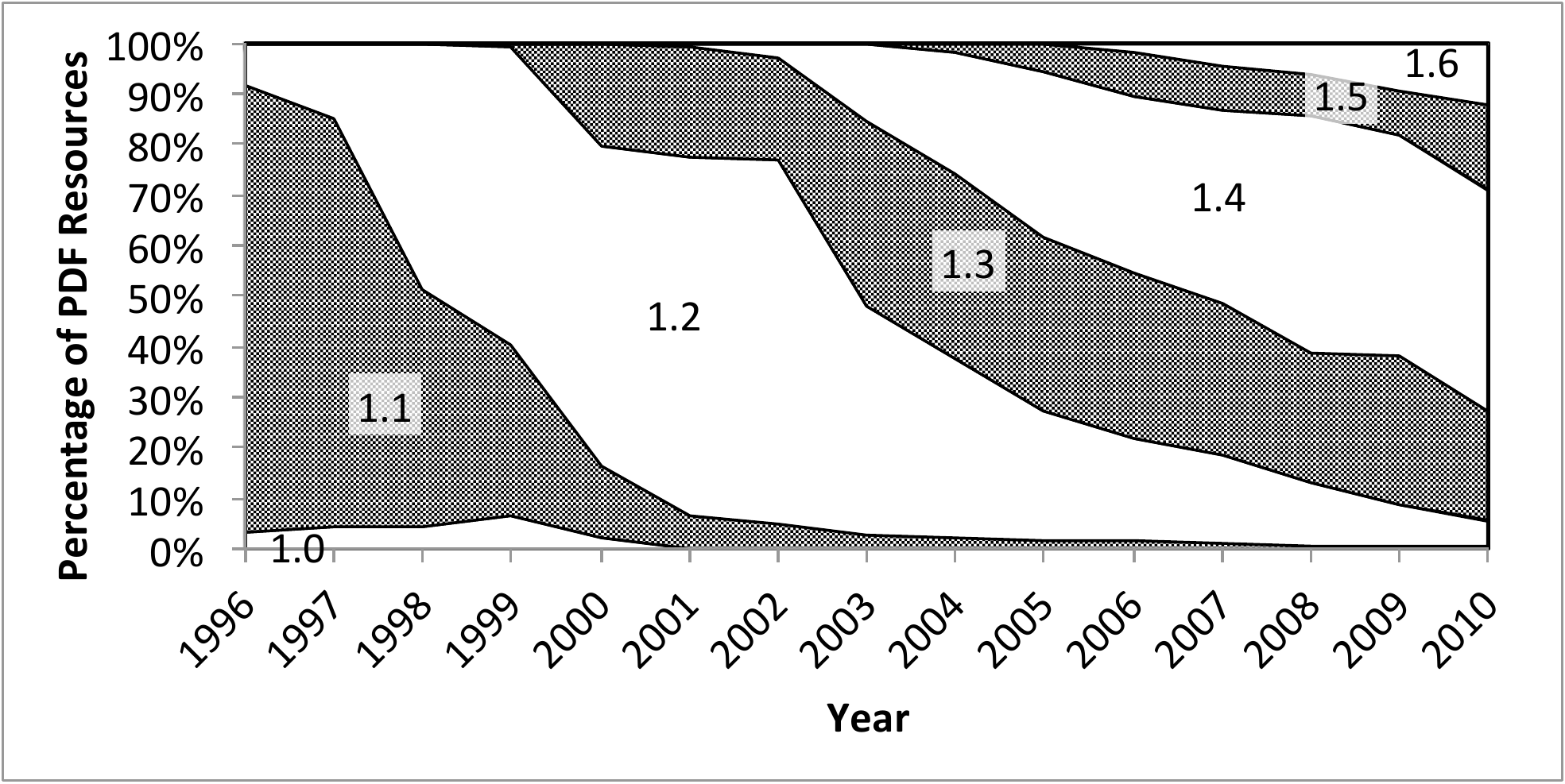}

\caption{\label{fig:Percentage-of-PDF}PDF versions over time. Formats identified
using DROID-B and Apache Tika.}
\end{figure}

\subsection{Versions \& Software}

Figures \ref{fig:Percentage-of-HTML} and \ref{fig:Percentage-of-PDF}
show how the popularity of various versions of HTML and PDF has changed
over time. In general, each new version grows and dominates the picture
for a few years, before very slowly sinking into obscurity. Thus,
while there were just two active versions of HTML in 1996 (2.0 and
3.2), all six were still active in 2010. Similarly, there were three
active versions of PDF in 1996 (1.0-1.2) and eleven different versions
in 2010 (1.0-1.7, 1.7 Extension Level 3, A-1a and A-1b, with 1.2-1.6
dominant). In general, it appears that format versions, like formats,
are quick to arise but slow to fade away.

\begin{figure}[t]
\includegraphics[width=1\columnwidth]{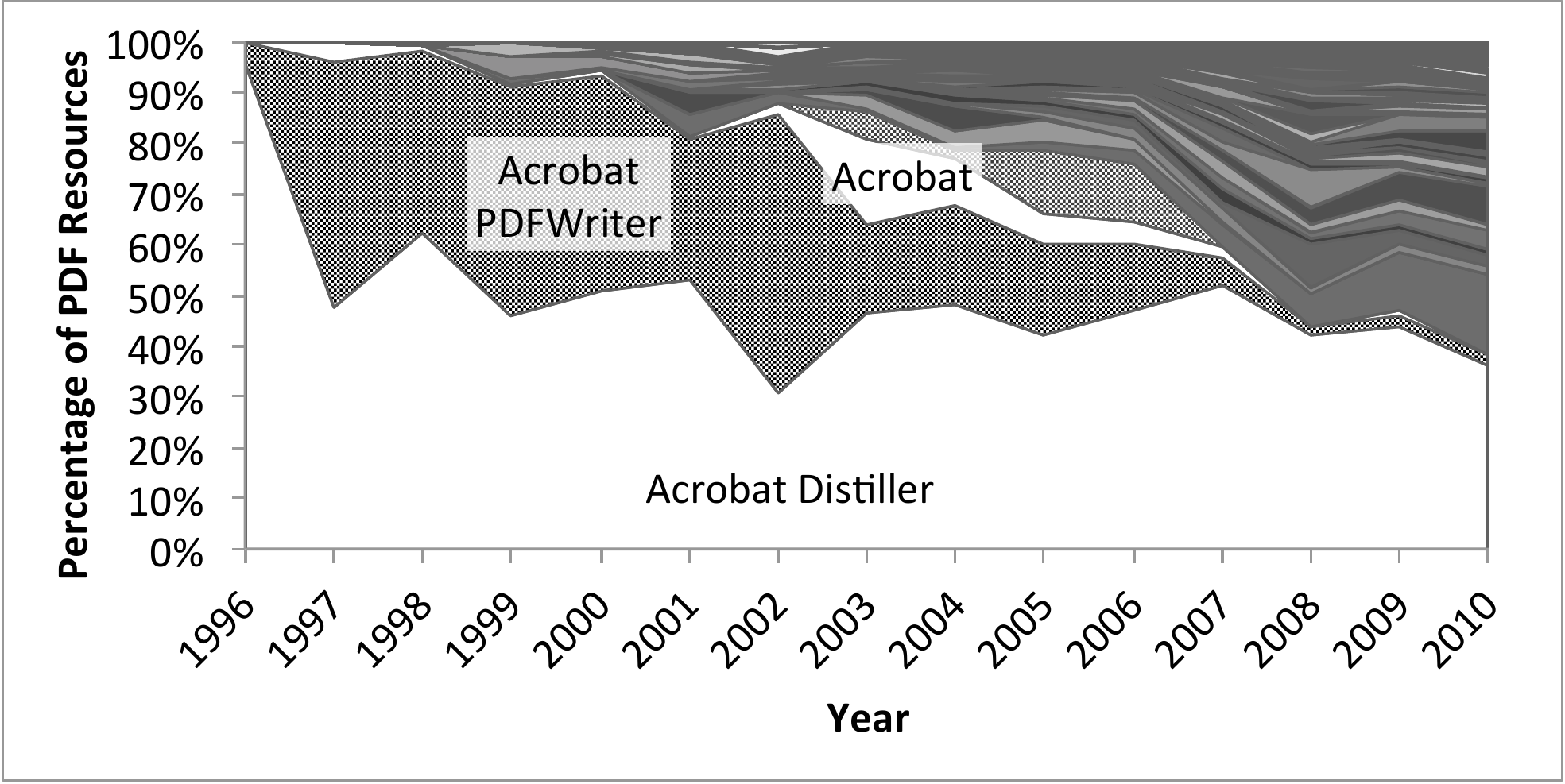}

\caption{\label{fig:Percentage-of-PDF-sw}PDF software identifiers over time.
Formats and software identified using Apache Tika.}
\end{figure}

Finally, figure \ref{fig:Percentage-of-PDF-sw} shows the popularity
of different software implementations over time and the dominance
of the Adobe implementations (although later years have seen an explosion
in the number of distinct creator applications, with over 2100 different
implementations of around 600 distinct software packages). Similarly,
the JPEG data revealed over 1900 distinct software identifiers and
over 2100 distinct hardware identifiers. We speculate that the number
of distinct implementations can be taken as an indicator for the maturity,
stability and degree of standardisation of a particular format, although
more thorough analysis across more formats would be required to confirm
this.

\section{Conclusions}

We have made a rich dataset available, profiling the format, version,
software and hardware data from large web archive spanning almost
one and a half decades. Our initial analysis supports Rosenthal's
position; that most formats last much longer than five years, that
network effects to appear to stabilise formats, and that new formats
appear at a modest, manageable rate. However, we have also found a
number of formats and versions that are fading from use, and these
should be studied closely in order to understand the process of obsolescence.
Furthermore, we must note that every corpus contains its own biases,
such as crawl size limits or scope parameters%
\footnote{Even the crawl time itself can be quite misleading, as a newly discovered
resource may have been created or published some years before%
}. Therefore, we recommend that similar analyses be performed on a
wider range of different corpora in order to attempt to confirm these
trends.

We used two different tools (DROID-B and Apache Tika) that perform
the essentially the same task (format identification), and ran them
across the same large and varied corpus. In effect, each can be considered
a different `opinion' on the format, and by uncovering the inconsistencies
and resolving them, we can improve the signatures and tools in a very
concrete and measurable way, and more rapidly approach something like
a `ground truth' corpus for format identification. 

Future work will examine whether the underlying biases of the corpus
can be addressed, whether we can reliably identify resources within
container formats, and whether the raw resource-level data can be
made available. This last point would allow many more format properties
to be exposed and make it easier to resolve inconsistent results by
linking back to the actual resources.

\section{Acknowledgments}

This work was partially supported by JISC (under grant DIINN06) and
by the SCAPE Project. The SCAPE project is co-funded by the European
Union under FP7 ICT-2009.4.1 (Grant Agreement number 270137).

\section{References}
\begin{lyxlist}{00.00.0000}
\item [{{[}1{]}}] Rothenberg, Jeff; (1997) \textquotedblleft{}Digital Information
Lasts Forever\textemdash{}Or Five Years, Whichever Comes First.\textquotedblright{}
RAND Video V-079
\item [{{[}2{]}}] Rothenberg, Jeff; (2012) ``Digital Preservation in Perspective:
How far have we come, and what's next?'' Future Perfect 2012, New
Zealand (Archived by WebCite\textregistered{} at \\
\href{http://www.webcitation.org/68OuQxEHj}{http://www.webcitation.org/68OuQxEHj})
\item [{{[}3{]}}] Rosenthal, David S.H.; (2010) \textquotedbl{}Format obsolescence:
assessing the threat and the defenses\textquotedbl{}, Library Hi Tech,
Vol. 28 Iss: 2, pp.195 - 210
\item [{{[}4{]}}] Clausen, Lars R.; (2004) ``Handling file formats''
(Archived by WebCite\textregistered{} at \\
\href{http://www.webcitation.org/68PyiaA9w}{http://www.webcitation.org/68PyiaA9w})
\item [{{[}5{]}}] Radtisch, Markus; May, Peter; Askov Blekinge, Asger;
Møldrup-Dalum, Per; (2012) ``SCAPE Deliverable D9.1: Characterisation
technology, Release 1 \& release report.'' (Archived by WebCite\textregistered{}
at \\
\href{http://www.webcitation.org/68OttmVnn}{http://www.webcitation.org/68OttmVnn}) \end{lyxlist}

\end{document}